\documentclass[letterpaper, 10 pt, conference]{ieeeconf}  

\IEEEoverridecommandlockouts
\usepackage[utf8]{inputenc}
\usepackage{authblk}

\title{\LARGE \bf
Causes of Misleading Statistics and Research Results Irreproducibility: A Concise Review
}

\author[1]{Farzan Shenavarmasouleh}
\author[1]{Hamid R. Arabnia}
\affil[1]{Department of Computer Science, University of Georgia, Athens, Georgia, United States, 30602 \authorcr {\{fs04199, hra\}@uga.edu}\vspace{1.5ex}}

\begin{document}
\maketitle
\thispagestyle{empty}
\pagestyle{empty}

\begin{abstract}
Bad statistics make research papers unreproducible and misleading. For the most part, the reasons for such misusage of numerical data have been found and addressed years ago by experts and proper practical solutions have been presented instead. Yet, we still see numerous instances of statistical fallacies in modern researches which without a doubt play a significant role in the research reproducibility crisis. In this paper, we review different bad practices that impact the research process from its beginning to its very end. Additionally, we briefly propose open science as a universal methodology that can facilitate the entire research life cycle.

\vspace{\baselineskip}

\end{abstract}
\section{\textbf{INTRODUCTION}}
For the past two centuries, the annual number of articles and journals have both kept growing at a steady rate of about 3\% and 3.5\% respectively. But, unsurprisingly, the increase in the number of researchers has accelerated this growth over the last few years. About 2.5 million articles get published each year worldwide \cite{ware2015stm} and yet, only a tiny proportion of their results are reproducible. In fact, more than 70\% of researchers claimed that they failed at reproducing another person's work and even more frighteningly, over 50\% could not reproduce their own work again, altogether leading to a reproducibility crisis \cite{baker20161}. 

One may assume that unproducible papers could only be found in poor-quality journals, but unfortunately, that is not the case. Many of these papers are written by very successful researchers and are published in top-gear journals of their fields. Given the fact that no good researcher and journal intend to present false or erroneous information to its audience, the question then arises as to why this incident happens. Knowing that the data do not speak for themselves and they have to be interpreted, it can be safely concluded that the main cause for this is statistics and it can affect the researches' outcomes at multiple levels.

Statistics is an essential part of every research, however, not much effort is being put into its proper education. Very few university majors require their students to take statistics courses and therefore many end up learning it via some available resources themselves. This could lead to a tremendous amount of problems, such as not using the correct statistical methods or graphs for a certain problem or make use of the ones which are not robust to the noises and outliers existing in the dataset. 

Sadly, there exists a bigger issue. Even if the paper itself is a hundred percent accurate, the words which get used to describe it can be fallacious. This could simply be unintentional and as before, purely due to bad education, but the numerical data can also be intentionally misused. Truth is not everyone's first priority. Thousands of news get published each day, many reporting a concern seeking to get resolved by politicians. Hence, they must compete in order to get noticed by the public and catch the attention of policymakers and if one succeeds, it consequently gets the biggest share of their time and budget while the others are left disregarded. As alarming news can bring more audiences to the corresponding media, and it means benefits, the people in charge do not hesitate to make use of such a thing. Even if it means that they have to falsely create them and statistics is the key component in this process. But, every move has consequences and propagandas like fake news can alter or even derail government programs and policies, leading to unforeseen circumstances.

This paper aims to address and classify the causes which make research papers unreproducible and explain how paper results could get reported misleadingly. We intend to look at these challenges as an opportunity to inform our audience. With this goal in mind, we hope that this manuscript would offer the readers, an insight and a better understanding of the causes of the problems; thus helping them not to become too reliant on papers that suffer from fallacies. We conclude the paper by proposing Open Science as an important addition to every research.
\section{\textbf{Resources and Tools}}
Since statistics is not a mandated part of many majors in universities, researchers in those fields mostly choose to learn it by themselves using some books and after learning the fundamentals, they move on to make use of certain software tools to do the complex and time-consuming calculations for them or generate eye-catching visuals to aid them in presenting the results of their work.
\subsection{\textbf{Textbooks}}
Bland et al. \cite{bland1988misleading} explain that most introductory statistic books that get published are written by authors who are experts in their fields but are rarely qualified to write about anything else. Those books tend to be more attractive to a large proportion of researchers of that certain field than a book authored by an experienced statistician, solely because the authors of those books share a common educational background with them and hence their words can be more comprehensible. 

From time to time, such books originate incorrect ideas. Bland argues that there were some cases in which the authors had made an error and then used a false argument to justify it. The horrifying part of this is that the people who read these books are unlikely to have enough knowledge to detect these flaws and consequently these errors keep propagating.
\subsection{\textbf{Software Tools}}
With the extensive use of computers, numerous software packages became accessible for research. Although they are massively useful and can save much time, using them without full understanding can be dangerous. Many statistics' operations have multiple versions, probably for different applications. For example, standard deviations could be calculated with n instead of n-1 in its denominator or it is incorrect to use unpooled variance instead of pooled variance in t-tests. These little differences could lead to getting significant results while there exists none \cite{bland1988misleading}.
\subsection{\textbf{Graphs}}
Graphs are considered great tools for visualizing massive complex data and their simplicity helps understand the contents better, but when it comes to statistics being misleading, graphs are the root of all evil. There are numerous types of graphs to choose from and a poor choice can result in a misleading visual. Much worse, they could easily be manipulated in a way to give the impression of a much better result.

Every misleading graph violates at least one of the followings:
\begin{itemize}
  \item A graph should always be equally spaced on all of its axes.
  \item If an axis has to show a quantity, it needs to start from zero. Otherwise, it fails at picturing the correct relative increase or decrease in the value.
  \item If using a pictograph, all symbols should be equally sized and the value which is being represented per symbol should be explicitly written.
\end{itemize}

\subsection{\textbf{Maps}}
Maps are used to illustrate spatial distributions of data such as cancer rates by county. However, if the sample size differs from area to area, the result can be highly deceptive especially in poorly-sampled areas. In addition, any adjustment such as using posterior means can cause further problems \cite{moulton1994potential} and therefore some spurious patterns could be found while in reality, they do not exist. Gelman et al. \cite{gelman1999all} state that although employing multiple imputed maps could help, they are still not appropriate for presenting the result as they may confuse the audience even more and hence they could not be generally used.
\section{\textbf{Data Collection}}
\subsection{\textbf{Broadened Problem Definition}}
The definition of the problem should always be specific and on point and this inevitably shrinks the size of the domain and decreases the resultant incident rates. It's common sense that bigger numbers are better and since these two are in conflict of interest with each other, it is not that rare to see definitions intentionally broadened to catch a bigger share of attention. For instance, in conducting a questionnaire for stranger abduction research, two different strategies could be used. One questionnaire could include short-term missings even if only for a few hours, attempted offenses which may not have even ended in actual abduction, to name a few, while the other only includes the children who went missing and found dead later on. The former results in around 15000 annual cases in the United States while the latter shows about 550 \cite{best1988missing}.

Back in the 90s, an exaggerated image of Australia's tourism was made showing it as a million-job industry, while in reality, the number of real jobs was around 200,000. In consequence, investors, business managers, politicians, and government were misled and by making wrong decisions, national efforts and resources were wasted and the job market was over-supplied. Leiper \cite{leiper1999conceptual} claims that this was partially because of the broad definition of the word tourist. Its definition from WTTC, WTO and many other institutions include all kinds of visitors who intend to stay somewhere for one or more nights with varying purposes, such as enjoying a holiday, pilgrim, visiting family and friends, doing business, going to school, staying at hospitals, etc. However, if the definition was constrained to a more fair one which only factored in the visitors who were there for leisure, then the numbers would have been closer to reality. 
\subsection{\textbf{Over-Discretized Sampling}}
Frequent sampling is a must when it comes to collecting helpful data. Nonetheless, when the time frames are longer than a standard threshold, the built dataset will fail to reveal valuable insights. As an example, hospital surge capacity, which is often measured by the number of empty beds that can get immediately ready to use in the case of an emergency, is usually reported annually and it fails to take the daily changes in the number of patients and the within-year changes in bed supply into account. When measured daily, the result shows way less availability \cite{delia2006annual}.
\subsection{\textbf{Biased Sampling}}
When you want to use a sample, you have to ensure that it can represent the larger population pretty well. The job of sample designs is to guarantee such behavior. If a design tends to favor a specific outcome, then it's considered a biased bad design. Biased samples can be created when individuals themselves choose to be involved in the research. The reason for this would be because the opinion of people who were not interested or simply didn't care enough has not been collected. Also, interviewing people only in specific places, such as streets, bars, libraries, and gyms or even same places in different cities can produce completely different results. To avoid biases, some good sampling methods have been developed, including but not limited to Simple Random Sampling (SRS), Stratified Random Sampling, Cluster Sampling, Systematic Sampling with Random Start, and Multistage Sampling \cite{madow1968elementary}.
\section{\textbf{Data Analysis and Statistical Methods}}
Gross domestic product (GDP) is one of the most popular indicators used to measure a country's economic health \cite{product2019second}. It depends on factors such as consumer spending, government spending, businesses' capital spending, and nation's total net exports while each of these, rely on many other elements related to the goods and services which the nation provides. Because of this innate complexity, calculation of GDP is an extremely hard and time-consuming process. In fact, too complex which even though nations' expert statisticians have the job of doing the calculations, from time to time, the results get revised multiple times afterward \cite{greenaway2014revisions}. As an example, the U.S. Bureau of Economic Analysis released three different GDP rates for the second quarter of 2015 \cite{gdp20151, gdp20152, gdp20153}.

The good news is not all statistics problems have that much complexity in terms of the numbers of factors present in them and can be calculated and analyzed quite simply. But, using the right methods is a must. Otherwise, the output can lead to misleading deductions. 

\subsection{\textbf{Common Scales}}
It's necessary to have a common scale in order to report and compare new results with baselines, and since there are usually multiple scales and methods to choose from, we should first fully understand the advantages and disadvantages of every single one of them and choose the best one for our work.

Results of clinical trials can be compared using several approaches. The most popular ones are Relative Risk (RR), Absolute Risk Reduction (ARR), Odds Ratio (OR), Log OR and Number Needed to Treat (NNT). It is common to use NNT as the default method because it's easier to comprehend but \cite{hutton2010misleading} refutes that this measurement is biased. NNT lacks precision as it rounds the result to its nearest integer which consequently can blur the differences among trials, it loses the time dimension and it has some fundamental issues such as the absence of a value that corresponds to no difference \cite{hutton2000number, lesaffre1999note, christensen2006number, stang2010common}. Hutton \cite{hutton2010misleading} further explains that ARR is a lot more reliable and should be used instead.
\subsection{\textbf{P-Value and Power}}
Technically speaking, P-value is the probability of getting a result, assuming the null hypothesis is true \cite{mellis2018lies}. Put in simple words, the null hypothesis is the theory that we want to reject, that is our experiment has no effect, and the alternative hypothesis is the opposite. After calculating the P, it is compared with a threshold, usually 0.05, and if it's above the cut line, the null hypothesis remains true. But, if P-value falls below the cut line, it means that either the null hypothesis is false, or although true, a very rare event has happened, both indicating that our experiment result is significant. 
There are some well-known downsides to using P-value. First, even though we know the smaller the P-value, the more significant the result is, it cannot be used to show the scale of the difference between null and alternative hypothesis. Second, if we fail to reject the null hypothesis, it does not indicate that our experiment has no effect whatsoever. It just tells us that there's not enough evidence that there is one. Thus, we can never be entirely sure. Third, the probability of the null hypothesis being true remains unclear as we have already assumed that it's true.

Besides these innate disadvantages, there exists a bigger problem. P-Hacking \cite{head2015extent} is the action of intentionally manipulating the P-value or the experiment to reach a significant result. The simplest form of it is to choose a bigger threshold than the resultant P and somehow justify it. Next is to divide the experiment into several smaller ones ($n$) and check all of them individually with the hope of finding a desirable P-value in at least one of them. This causes family-wise error and the problem is as the P threshold ($\alpha$) remains untouched and the same as the original experiment in all of them, the probability of getting at least one significant result just by chance would be $1-(1-\alpha)^n$. Then usually, only the significant experiment gets published without providing any context of the bigger picture which is clearly misleading and there's a huge chance that the significance is only the result of a Type I error and hence, most probably it cannot be reproduced. Bonferroni correction has to be used in order to avoid family-wise error and it suggests to use $\frac{P}{n}$ as the new threshold for each of the smaller experiments \cite{weisstein2004bonferroni}.

In this context, type II error means failing to reject the null hypothesis when it's false.  While $\alpha$ corresponds to getting a false positive or being a victim of type I error, $\beta$ is the probability of Type II error or getting a false negative. Power is the probability of detecting an effect when there exists one and its value is $1-\beta$. In a certain problem $\alpha + \beta$ equals to a fixed number. So, it can be concluded that there's a trade-off between them and depending on the problem, the researcher have to decide which error should be valued more. Increasing sample size can do good to both of them as it makes the distributions more accurate and thinner and as a result reduces the possibility of both errors, by making that fixed number smaller and resulting in a bigger power and a more precise P-value.
\subsection{\textbf{Robust Statistics}}
Classical statistics, also called parametric statistics, require data to be normally distributed. Even though there is no guarantee that the data in smalls samples are normally distributed, more than often it is assumed that they are and the calculations are done based on that. One alternative is to use non-parametric statistics that do not have this requirement, but they have their own set of rules as well. Using parametric statistics on not normally distributed data and using non-parametric statistics on normally distributed data both result in less power \cite{larson2012our}. As aforementioned, one way to increase the power, which happens to be the hardest way as well, is to increase the sample size. One alternative is to value Type II error more and by raising $\alpha$ make the $\beta$ smaller and as a result get a higher power. Second alternative is to use a more powerful statistics, namely robust statistics \cite{wilcox2011introduction}. It proposes principled ways to overcome parametric and non-parametric issues. For instance, it suggests to use effect size and confidence intervals in addition to P-value since unlike the latter, the result of those stay the same if $\alpha$ changes \cite{sullivan2012using, mellis2018lies}.

Outliers can almost be found in every population. If ignored, they can violate the assumption of normal distribution and result in less power, and if removed manually they can cause non-Independence in remaining data, and in some cases, they could be hard to be found in the first place. One solution that robust statistics proposed is to trim both ends and readjust the normal equation to compromise the effect of non-independence. It also recommends ways to do t-tests, correlations, 1-way ANOVA, etc. in a robust manner \cite{larson2012our}. It's worth mentioning that robust statistics might result in a different outcome in terms of significance and non-significance of the research from the conventional statistics but it's most certainly more accurate.
\section{\textbf{Fallacious Deductions}}
Statistics is a subset of mathematics and it should not come as a surprise that understanding, mapping and modeling the concepts may not be a straightforward process. Even experts can find contradictory results with each other and at times such as in the case of Sally Clark \cite{nobles2005misleading}, the deductions that have made based on those results could be a matter of life or death. 

Misinterpretations can affect future research, resources, and fundings as well. For instance, while analyzing the bullets spread on returned American planes at World War II, it has been noticed that most bullets were around the fuselage and least around the engines. The initial reasoning was that in order to increase the survival rate, more armor should be placed on the fuselage area. Wald \cite{wald1980reprint} refuted that conclusion and explained that the reason most planes had more bullet hits on their fuselage was that the ones which got hit on their engines could not eventually make it home and as a result the distribution was uneven. Thus, to increase the chance of survival, more armor is needed on the engine surfaces.
\subsection{\textbf{Confusing Correlation With causality}}
Scatter plots and covariance matrices are used to find relationships between pairs of features in the data. If their relationship is linear, Pearson's correlation ($-1 \leq r \leq 1$) \cite{sedgwick2012pearson} can be useful and just by looking at its sign and value, we can find out that whether the correlation is positive or inverse and how strong it is. If a strong correlation is found, it could mean two things. First, X causes Y or Y causes X. Second, either there is another reason Z that causes both X and Y or they are related by complete coincidence. The latter is called spurious correlation \cite{simon1954spurious} and means even though these two features seem to correlate with each other, there's no way one can cause another and this confusion between correlation and causality can lead to erroneous conclusions. Besides, not all correlations are linear and therefore scatter plots should always be analyzed.
\subsection{\textbf{Ecological Fallacy}}
Ecological fallacy happens if a deduction about an individuals' characteristic is made based on the results found for a group in which that person belongs to \cite{piantadosi1988ecological, freedman1999ecological}. Data are collected at different levels such as a continent, a country, or a state. As previously discussed, the features can be analyzed in terms of correlation and then lead to discoveries. However, these findings are only valid in the level in which they are being analyzed and if any deduction is made based on them for lower-level groups or individuals, it cannot be trusted.
\subsection{\textbf{Will Rogers Phenomenon}}
Assume that there are two groups and all the people in group A have lower IQ than all the people in group B. If the dumbest person in group B moves to group A, the average IQ for both groups rises since B will be got rid of its least intelligent person who was dragging the average down and A will profit from gaining its smartest member. This effect is called the Will Rogers phenomenon and has caused some erroneous conclusions especially in the area of cancer screening \cite{feinstein1985will,gofrit2008will,albertsen2005prostate}. Feinstein \cite{feinstein1985will} elaborates the very first instance of this phenomenon by showing that improvements reported in the rates of survival of lung cancer patients are flawed. He explains that patients were originally divided into two groups. The ones who their cancer was localized and the ones with higher stages of cancer and metastasis. With the advancements in the technology, micro-metastases could be found in people from the localized group moving those patients to the other group. This simply improved the survival rates in both groups since with this migration, the first group has lost its patients with lower life expectancy and the second group earned a few patients with better overall health than its usual patients.
\section{\textbf{Indecent Reporting}}
When reporting the result, it's conventional to use percentages, but they can be highly misleading as they do not give any information about the initial and end value or at least the difference that has been made. A bigger picture must always be clearly drawn to better understand how significant the new result is. Also, the results should never and under no circumstances, get rounded to their nearest bigger number just to attract more attention. Another bad practice is hiding or not providing enough context to the audience, in order to mislead or fool them into believing that the results were significant. An example of this as aforementioned, could be not using Bonferroni correction and not talking about the original experiment at the same time. Sowell \cite{sowell2011economic} also argues about a few cases in household reports. Such as not mentioning the change that has occurred in the number of people in families while comparing it to old statistics. Assume that in the past each household had 6 people in it and at a later year this number is decreased to 4. Even if the income per person is increased by 25 percent, the total sum shows an economic decline since it equals the amount that 5 people could make in past.
\section{\textbf{Open Science as a possible solution}}
Open Science is a group of ideas and principles promoting openness and transparency of the research in every stage of its life cycle while keeping the integrity of the researcher's work. It starts with sharing the main idea with the intention of inducing collaboration with other researchers, funders, publishers, industries and institutions and then quickly advancing the research to its final stage. Every note, data, code, software and the end result including the paper should be publicly available to everyone and ready to be reused. The long term goal of open science is to encourage good and high-quality research regardless of its outcome by valuing the research process more than its result and judge researchers based on their work and contribution to the community instead of completely relying on formal publications. Open science allows and even inspires researchers to replicate other researches and also publish the negative results which it believes can be very informative. This acts against the public and perish culture \cite{bedeian2010management} which motivates researchers to use biases, P-hacking and all the other bad practices mentioned in this paper to advance their career and instead enable researchers to do researches based on their joy and curiosity \cite{molloy2011open,nosek2015promoting, woelfle2011open}.

\section{\textbf{Conclusion and Discussion}}
A tremendous amount of data resides in every research. It's almost impossible to analyze and find something meaningful from that data without using statistics. However, if used incorrectly, either intentionally or unintentionally, it can result in erroneous and misleading conclusions and thus make the research unreproducible. It is important to know the common causes of this incident to avoid it. In this review, we attempted to summarize the most common flaws that can happen while using statistics and although it's unfeasible to describe each bad practice comprehensively, we believe, all in all, it gives a rough overview of them and how they could be avoided.


\section{\textbf{Conflict of Interest}}
The authors declare that there is no conflict of interest regarding the publication of this article.

\addtolength{\textheight}{-12cm}

\bibliography{bib} 

\begin{thebibliography}{10}

\bibitem{ware2015stm}
M.~Ware and M.~Mabe, ``The stm report: An overview of scientific and scholarly
  journal publishing,'' 2015.

\bibitem{baker20161}
M.~Baker, ``1,500 scientists lift the lid on reproducibility,'' {\em Nature
  News}, vol.~533, no.~7604, p.~452, 2016.

\bibitem{bland1988misleading}
J.~M. BLAND and D.~G. ALTMAN, ``Misleading statistics: errors in textbooks,
  software and manuals,'' {\em International journal of epidemiology}, vol.~17,
  no.~2, pp.~245--247, 1988.

\bibitem{moulton1994potential}
L.~H. Moulton, B.~Foxman, R.~A. Wolfe, and F.~K. Port, ``Potential pitfalls in
  interpreting maps of stabilized rates,'' {\em Epidemiology}, pp.~297--301,
  1994.

\bibitem{gelman1999all}
A.~Gelman and P.~N. Price, ``All maps of parameter estimates are misleading,''
  {\em Statistics in medicine}, vol.~18, no.~23, pp.~3221--3234, 1999.

\bibitem{best1988missing}
J.~Best, ``Missing children, misleading statistics,'' {\em The Public
  Interest}, vol.~92, p.~84, 1988.

\bibitem{leiper1999conceptual}
N.~Leiper, ``A conceptual analysis of tourism-supported employment which
  reduces the incidence of exaggerated, misleading statistics about jobs,''
  {\em Tourism Management}, vol.~20, no.~5, pp.~605--613, 1999.

\bibitem{delia2006annual}
D.~DeLia, ``Annual bed statistics give a misleading picture of hospital surge
  capacity,'' {\em Annals of Emergency Medicine}, vol.~48, no.~4, pp.~384--388,
  2006.

\bibitem{madow1968elementary}
W.~G. Madow, ``Elementary sampling theory,'' 1968.

\bibitem{product2019second}
``Gross domestic product, second quarter 2019 (third estimate),'' {\em U.S.
  Bureau of Economic Analysis news release}, September 26, 2019.

\bibitem{greenaway2014revisions}
R.~Greenaway-McGrevy, B.~Grimm, and D.~Fixler, ``The revisions to gdp, gdi, and
  their major components,'' 2014.

\bibitem{gdp20151}
``Gross domestic product, 2nd quarter 2015 (advance estimate),'' {\em U.S.
  Bureau of Economic Analysis news release}, July 30, 2015.

\bibitem{gdp20152}
``Gross domestic product, 2nd quarter 2015 (second estimate),'' {\em U.S.
  Bureau of Economic Analysis news release}, August 27, 2015.

\bibitem{gdp20153}
``Gross domestic product, 2nd quarter 2015 (third estimate),'' {\em U.S. Bureau
  of Economic Analysis news release}, September 25, 2015.

\bibitem{hutton2010misleading}
J.~L. Hutton, ``Misleading statistics,'' {\em Pharmaceutical Medicine},
  vol.~24, no.~3, pp.~145--149, 2010.

\bibitem{hutton2000number}
J.~Hutton, ``Number needed to treat: properties and problems,'' {\em Journal of
  the Royal Statistical Society: Series A (Statistics in Society)}, vol.~163,
  no.~3, pp.~381--402, 2000.

\bibitem{lesaffre1999note}
E.~Lesaffre and G.~Pledger, ``A note on the number needed to treat,'' {\em
  Controlled Clinical Trials}, vol.~20, no.~5, pp.~439--447, 1999.

\bibitem{christensen2006number}
P.~M. Christensen and I.~S. Kristiansen, ``Number-needed-to-treat (nnt)--needs
  treatment with care,'' {\em Basic \& clinical pharmacology \& toxicology},
  vol.~99, no.~1, pp.~12--16, 2006.

\bibitem{stang2010common}
A.~Stang, C.~Poole, and R.~Bender, ``Common problems related to the use of
  number needed to treat,'' {\em Journal of clinical epidemiology}, vol.~63,
  no.~8, pp.~820--825, 2010.

\bibitem{mellis2018lies}
C.~Mellis, ``Lies, damned lies and statistics: clinical importance versus
  statistical significance in research,'' {\em Paediatric respiratory reviews},
  vol.~25, pp.~88--93, 2018.

\bibitem{head2015extent}
M.~L. Head, L.~Holman, R.~Lanfear, A.~T. Kahn, and M.~D. Jennions, ``The extent
  and consequences of p-hacking in science,'' {\em PLoS biology}, vol.~13,
  no.~3, p.~e1002106, 2015.

\bibitem{weisstein2004bonferroni}
E.~W. Weisstein, ``Bonferroni correction,'' 2004.

\bibitem{larson2012our}
J.~Larson-Hall, ``Our statistical intuitions may be misleading us: Why we need
  robust statistics,'' {\em Language Teaching}, vol.~45, no.~4, pp.~460--474,
  2012.

\bibitem{wilcox2011introduction}
R.~R. Wilcox, {\em Introduction to robust estimation and hypothesis testing}.
\newblock Academic press, 2011.

\bibitem{sullivan2012using}
G.~M. Sullivan and R.~Feinn, ``Using effect size—or why the p value is not
  enough,'' {\em Journal of graduate medical education}, vol.~4, no.~3,
  pp.~279--282, 2012.

\bibitem{nobles2005misleading}
R.~Nobles and D.~Schiff, ``Misleading statistics within criminal trials: the
  sally clark case,'' {\em Significance}, vol.~2, no.~1, pp.~17--19, 2005.

\bibitem{wald1980reprint}
A.~Wald, ``A reprint of'a method of estimating plane vulnerability based on
  damage of survivors.,'' tech. rep., CENTER FOR NAVAL ANALYSES ALEXANDRIA VA
  OPERATIONS EVALUATION GROUP, 1980.

\bibitem{sedgwick2012pearson}
P.~Sedgwick, ``Pearson’s correlation coefficient,'' {\em Bmj}, vol.~345,
  p.~e4483, 2012.

\bibitem{simon1954spurious}
H.~A. Simon, ``Spurious correlation: A causal interpretation,'' {\em Journal of
  the American statistical Association}, vol.~49, no.~267, pp.~467--479, 1954.

\bibitem{piantadosi1988ecological}
S.~Piantadosi, D.~P. Byar, and S.~B. Green, ``The ecological fallacy,'' {\em
  American journal of epidemiology}, vol.~127, no.~5, pp.~893--904, 1988.

\bibitem{freedman1999ecological}
D.~A. Freedman, ``Ecological inference and the ecological fallacy,'' {\em
  International Encyclopedia of the social \& Behavioral sciences}, vol.~6,
  no.~4027-4030, pp.~1--7, 1999.

\bibitem{feinstein1985will}
A.~R. Feinstein, D.~M. Sosin, and C.~K. Wells, ``The will rogers phenomenon:
  stage migration and new diagnostic techniques as a source of misleading
  statistics for survival in cancer,'' {\em New England Journal of Medicine},
  vol.~312, no.~25, pp.~1604--1608, 1985.

\bibitem{gofrit2008will}
O.~N. Gofrit, K.~C. Zorn, G.~D. Steinberg, G.~P. Zagaja, and A.~L. Shalhav,
  ``The will rogers phenomenon in urological oncology,'' {\em The Journal of
  urology}, vol.~179, no.~1, pp.~28--33, 2008.

\bibitem{albertsen2005prostate}
P.~C. Albertsen, J.~A. Hanley, G.~H. Barrows, D.~F. Penson, P.~D. Kowalczyk,
  M.~M. Sanders, and J.~Fine, ``Prostate cancer and the will rogers
  phenomenon,'' {\em Journal of the National Cancer Institute}, vol.~97,
  no.~17, pp.~1248--1253, 2005.

\bibitem{sowell2011economic}
T.~Sowell, {\em Economic facts and fallacies}.
\newblock Basic Books, 2011.

\bibitem{bedeian2010management}
A.~G. Bedeian, S.~G. Taylor, and A.~N. Miller, ``Management science on the
  credibility bubble: Cardinal sins and various misdemeanors,'' {\em Academy of
  Management Learning \& Education}, vol.~9, no.~4, pp.~715--725, 2010.

\bibitem{molloy2011open}
J.~C. Molloy, ``The open knowledge foundation: open data means better
  science,'' {\em PLoS biology}, vol.~9, no.~12, p.~e1001195, 2011.

\bibitem{nosek2015promoting}
B.~A. Nosek, G.~Alter, G.~C. Banks, D.~Borsboom, S.~D. Bowman, S.~J. Breckler,
  S.~Buck, C.~D. Chambers, G.~Chin, G.~Christensen, {\em et~al.}, ``Promoting
  an open research culture,'' {\em Science}, vol.~348, no.~6242,
  pp.~1422--1425, 2015.

\bibitem{woelfle2011open}
M.~Woelfle, P.~Olliaro, and M.~H. Todd, ``Open science is a research
  accelerator,'' {\em Nature Chemistry}, vol.~3, no.~10, p.~745, 2011.

\end{thebibliography}
\bibliographystyle{ieeetr}

\end{document}